# Fragile systems: A hidden-variable Bayesian framework leading to quantum theory

Yasmín Navarrete [a,*], Sergio Davis [b,c]

[a] Instituto de Filosofía y Ciencias de la Complejidad, Los Alerces 3024 Ñuñoa, Santiago, Chile
[b] Research Center in the Intersection of Plasma Physics, Matter and Complexity ($P^2mc$), Comisión Chilena de Energía Nuclear, Casilla 188-D, Santiago, Chile
[c] Departamento de Física, Facultad de Ciencias Exactas, Universidad Andres Bello, Sazie 2212, Santiago, Chile

ARTICLE INFO

ABSTRACT

An understanding of quantum theory in terms of new, underlying descriptions capable of explaining the existence of non-classical correlations, non-commutativity of measurements and other unique and counter-intuitive phenomena remains still a challenge at the foundations of our description of physical phenomena. Among some proposals, the idea that quantum states are essentially states of knowledge in a Bayesian framework is an intriguing possibility due to its explanatory power. In this work, the formalism of quantum theory is derived from the application of Bayesian probability theory to "fragile" systems, that is, systems that are perturbed by the measurement. Complex Hilbert spaces, non-commuting operators and the trace rule for expectations all arise naturally from the use of linear algebra to solve integral equations involving classical probabilities over hidden variables. The non-fragile limit of the theory, where all measurements are commutative and the theory becomes analogous to classical statistical theory is discussed as well.

**Contents**

## Introduction

The meaning and the interpretation of quantum theory has been a controversial issue in physics because it suggests a non-local nature of our universe. In their seminal paper in 1935, Einstein, Podolsky and Rosen [1](EPR) asserted that quantum theory was incomplete, because of the impossibility of predicting complementary quantities such as position and velocity of a particle at the same time. Einstein, in the EPR paper [2], argued that this uncertainty in momentum should not be a problem if quantum mechanics provided a complete description of "reality". He believed in the existence of "elements of reality", which were properties of particles that had definite values, even if not all of these values could be simultaneously measured (as per Heisenberg's

* Corresponding author.
*E-mail addresses:* yasmin.navarrete@gmail.com, y.navarretediaz@uandresbello.edu (Y. Navarrete).

Uncertainty Principle). So, he concluded that there must be hidden variables (undetermined parameters) governing particle behavior that, if known, would allow for a complete description of their states.

Moreover, quantum entanglement introduced the notion of non-locality, which was later discussed by means of Bell's theorem [2]. Entangled systems manifest non-classical correlations between outcomes performed on physical systems that are far apart, but such that they have interacted in the past. It is a fundamental phenomenon in quantum mechanics and it has become a central feature of this field. Entanglement is sometimes described as "spooky action at a distance" due to its non-intuitive and counterintuitive nature. In other words, the properties of entangled particles are interconnected, even when they are separated by vast distances.

When experiments demonstrate violations of Bell's inequalities, it implies that the observed correlations between entangled particles (such as in the case of quantum entanglement experiments) cannot be explained by any theory that maintains both locality and realism. This leads to the conclusion that quantum mechanics, if to be understood in terms of hidden variables (additional parameters that determine the outcomes of quantum measurements), cannot be a local theory.

As a consequence of experimental violations of Bell's inequality [3] it was concluded that quantum theory, where correlations between entangled particles cannot be explained by any theory that maintains both locality and realism, is therefore to be formulated in terms of hidden variables, and has to be a non-local theory. In order to attempt to fill this void in the understanding of the foundations of quantum theory, a number of hidden-variables theories [2,4–7] have been proposed.

Furthermore, recently there have been interest in successful applications of the mathematical formalism of quantum theory in widely different contexts, both in Physics and outside of it. For instance, in the fields of quantum cognition [8], machine learning [9], signal processing [10], classical hydrodynamical systems [11,12]. Despite their success, a formal theoretical justification of the application of quantum theory to those systems outside of the traditional quantum realm is still lacking.

In this work, we will formulate a theory of *fragile systems* [13] (systems which are modified by the measurement) based on hidden variables and derived from the application of Bayesian probability. We will recover the formalism of quantum theory from first principles, in particular, we will obtain that:

- The states after a measurement correspond to invariants of a linear transformation.
- This transformation leads to an eigenvalue equation involving a linear operator in Hilbert space.
- Expectations are given by the trace rule of the density matrix formalism.

## Fragile systems

In simple terms, a fragile system is one that is affected by measurement performed on it. This distinguishes it from a non-fragile (classical) system, which is not modified upon measurement (we will think of a measurement as an interaction between two systems where one of them acquires information about the other).

Because any system (being fragile or not) possesses information, we will think of a system as a "black box" that can be found in different **internal states**, to be denoted by $\lambda$. In general $\lambda$ contains many degrees of freedom, but we will not make use of that inner structure here. The internal state $\lambda$ (which changes whenever a measurement is performed on the system) contains all the information necessary to describe the system.

We will consider a system with several real-valued, discrete observables $A, B, C, \ldots$. To each observable $x$ we will associate a real function $R_x$. For instance, the observable $A$ may yield a value given by a real function $R_A(\lambda) \in \{a_1, \ldots, a_N\}$. In this case, the statement $a_k = R_A(\lambda)$ means that a measurement of $A$ when the system is found in its internal state $\lambda$ produced the value $a_k$.

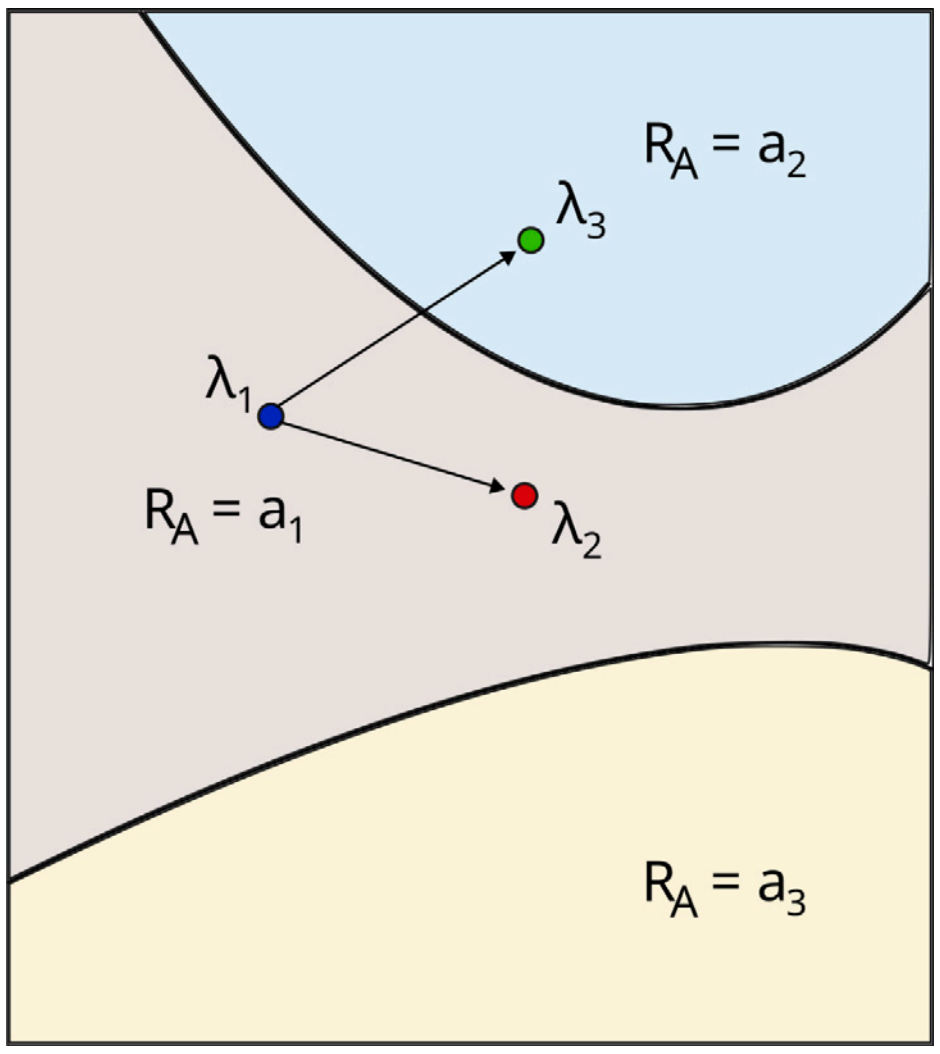

**Fig. 1.** Hidden variables and the function $R_A(\lambda)$: Each region corresponds to a value of an observable related with a distribution of hidden variables.

The crucial difference between a fragile system and a non-fragile one is that, in a fragile system, **access to the internal state $\lambda$ is impossible,** because it is precisely this internal state which is modified by the measurement. We cannot, therefore, assume that we can evaluate in practice $R_A$ on the internal state $\lambda$ to obtain the outcome of the measurement since every measurement outcome has associated a region of possible state $\lambda$ as it is depicted in Fig. 1

As the modification of the state $\lambda$ depends on the details of the environment doing the measurement (which we do not know or control with accuracy), the outcome of measurement is unavoidably stochastic, and a mathematical formulation requires probability theory.

In summary, a fragile system is one that (a) gets modified upon a measurement (b) when it is measured, the system remains in one "macroscopic" state and (c) its measurable properties have finite outcomes.

## Probability theory

As we do not have exact knowledge of the internal state $\lambda$, we can only assign a probability distribution over it, let us say $P(\lambda|S)$, in **our state of knowledge** $S$. Unlike non-fragile systems, in a fragile system there is no state of knowledge $\mathcal{I}$ corresponding to an infinitely sharp peak, $P(\lambda|\mathcal{J}) = \delta(\lambda - \lambda_0)$. Neither can we know the exact modification that a measurement will do on the internal state $\lambda$, thus for an observable $A$ we can only assign a transition probability $P(\lambda'|\lambda, A)$ of the final internal state $\lambda'$ given the initial state $\lambda$ and that a measurement of $A$ has occurred.

By the application of the marginalization rule of probability [14], we see that if we are in a state of knowledge $S$ before a measurement of $A$ is made, after the measurement the new state of knowledge $S'$ will be given by

$$P(\lambda'|S') = \int d\lambda P(\lambda|S)P(\lambda'|\lambda, A) = P(\lambda'|S, A), \tag{1}$$

so that $S' = S \wedge A$ (that is the previous state S and the fact that $A$ has been measured). In the particular case of a non-fragile system, the internal state $\lambda$ is not modified by the measurement of $A$, and therefore $P(\lambda'|\lambda, A) = \delta(\lambda' - \lambda)$, and we have thus $P(\lambda|S') = P(\lambda|S)$.

We are going to consider the situation after a measurement of $A$ yields the value $a_k$, i.e. when we have fixed points of this setting. Bayes' theorem [14] tells us that our state of knowledge must agree with the probability rules,

$$P(\lambda|a_k) = \frac{P(a_k|\lambda)P(\lambda|\mathcal{I}_0)}{P(a_k|\mathcal{I}_0)}, \tag{2}$$

wherein $\mathcal{I}_0$ is the initial state of knowledge. So we have

$$P(a_k|\mathcal{I}_0) = \int d\lambda P(\lambda|\mathcal{I}_0)\delta(R_A(\lambda), a_k). \tag{3}$$

This implies our state is one of complete knowledge of $R_A$. We will postulate that the prior probability of the internal states $P(\lambda|\mathcal{I}_0)$ is flat, and then Eq. (2) reduces to

$$P(\lambda|a_k) = \frac{\delta(R_A(\lambda), a_k)}{\Omega_A(a_k)} = \begin{cases} \frac{1}{\Omega_A(a_k)} & \text{if } R_A(\lambda) = a_k, \\ 0 & \text{if } R_A(\lambda) \neq a_k. \end{cases} \tag{4}$$

where $\Omega_A(a_k) = \int d\lambda \delta(R_A(\lambda), a_k)$ is the density of internal states with given value of $R_A$. The fact that $P(\lambda|a_k)$ forbids all values of $\lambda$ with $R_A \neq a_k$ implies that two consecutive measurements of the same observable $A$, without any perturbation in between, will yield the same outcome $a_k$. From this it follows that the state of knowledge after a measurement must be an invariant (i.e. an "eigenfunction") of the transformation $S \to S'$ given by Eq. (1). That is, if we define $g_k(\lambda) := P(\lambda|a_k)$ as the state of knowledge, after obtaining the outcome $a_k$ in the measurement of $A$, we must have that $g_k$ is unaffected by the transformation, so

$$g_k(\lambda') = \int d\lambda g_k(\lambda)P(\lambda'|\lambda, A). \tag{5}$$

Obviously, Eq. (5) is automatically true for all $g_k(\lambda)$ in a non-fragile system as $P(\lambda'|\lambda, A) = \delta(\lambda' - \lambda)$. In general there is only a finite set of functions $g_k$ that are solutions of Eq. (5).

## Representation in terms of a complete basis

In this section we introduce the use of a notation in terms of linear algebra, in order to bring the notation in the natural language of quantum theory. We can construct a complete, orthonormal basis $\{\phi_1(\lambda), \ldots, \phi_N(\lambda)\}$ for the probabilities $P(\lambda|S)$, as follows. We introduce a new observable $E$ with possible outcomes $e_1, \ldots, e_N$. Then, using the marginalization rule and replacing $P(\lambda|e_k)$ according to Eq. (4) for $E$ instead of $A$, we have,

$$P(\lambda|S) = \sum_{k=1}^{N} P(\lambda|e_k)P(e_k|S) = \sum_{k=1}^{N} \frac{\delta(R_E(\lambda), e_k)}{\Omega_E(e_k)} P(e_k|S). \tag{6}$$

Now let us define the basis functions

$$\phi_i(\lambda) := \frac{\delta(R_E(\lambda), e_i)}{\sqrt{\Omega_E(e_i)}}, \tag{7}$$

although this is not the only possible choice: we can propose different orthonormal functions (e.g. complex functions). Taking up the formulation given in Eq. (7) $P(\lambda|S)$ is written in this basis with coefficients $v_i$, i.e.,

$$P(\lambda|S) = \sum_{i=1}^{N} v_i \phi_i(\lambda). \tag{8}$$

This fixes the coefficients $v_i = P(e_i|S)/\sqrt{\Omega_E(e_i)}$ according to Eq (6). Because the function $R_E(\lambda)$ is single-valued, $\phi_i(\lambda)\phi_j(\lambda) = 0$ for any $\lambda$ if $i \neq j$. Furthermore, $\phi_i(\lambda)^2 = P(\lambda|e_i)$, so the basis is orthonormal, that is,

$$\int d\lambda \phi_i(\lambda)\phi_j(\lambda) = \delta_{ij}. \tag{9}$$

Expanding also $P(\lambda'|S')$ in terms of this basis as

$$P(\lambda'|S') = \sum_{j=1}^{N} w_j \phi_j(\lambda'), \tag{10}$$

we can represent the states of knowledge $S$ and $S'$ by the vectors $\boldsymbol{v} = (v_1, \ldots, v_N)$ and $\boldsymbol{w} = (w_1, \ldots, w_N)$, respectively. We show in the appendix A that Eq. (1) is equivalent to the matrix equation

$$\boldsymbol{w} = \mathbb{T}_A \cdot \boldsymbol{v}, \tag{11}$$

with components

$$w_k = \sum_{i=1}^{N} T_{ki}^{(A)} v_i, \tag{12}$$

where we have defined the matrix $\mathbb{T}_A$ with elements

$$T_{ij}^{(A)} = \int d\lambda d\lambda' \phi_i(\lambda')P(\lambda'|\lambda, A)\phi_j(\lambda). \tag{13}$$

The fixed points of the transformation, namely the functions $g_k(\lambda)$, are now encoded as the eigenvectors $\boldsymbol{u}_k$ (with eigenvalue 1) such that $\boldsymbol{u}_k = \mathbb{T}_A \cdot \boldsymbol{u}_k$. The matrix $\mathbb{T}_A$ is the transformation which allows us to obtain the fixed points of the system for the observable $A$. On the other hand, we see that we can obtain an analogous operator $\mathbb{A}$ leading to the necessary transformation to obtain also the outcomes of the measurement as its eigenvalues. The matrix elements $A_{ij}$ are given by (see Section Appendix B "The eigenvalue equation for the matrix $\mathbb{A}$"),

$$A_{ij} = \int d\lambda d\lambda' R_A(\lambda)\phi_i(\lambda')P(\lambda'|\lambda, A)\phi_j(\lambda), \tag{14}$$

which then allow us to write the eigenvalue problem as

$$a_i \boldsymbol{u}_i = \mathbb{A} \cdot \boldsymbol{u}_i. \tag{15}$$

The matrix elements $A_{ij}$ are real numbers, because the function $R_A(\lambda)$ and the basis functions $\{\phi_i(\lambda)\}$ are real. However, in general it can be more convenient to express the eigenvalue problem in an arbitrary complex basis $\{\psi_i(\lambda)\}$ since a real matrix can have complex eigenvectors: complex eigenvectors typically appear in conjunction with complex eigenvalues [15], so that

$$\mathbb{A} = \sum_{i=1}^{N} a_i \boldsymbol{u}_i \boldsymbol{u}_i^T \to \sum_{i=1}^{N} a_i \boldsymbol{c}_i \boldsymbol{c}_i^{*T}. \tag{16}$$

with $\boldsymbol{c}_k$ a complex vector of dimension $N$, namely the coefficients of $\phi_k(\lambda)$ in the complex basis $\{\psi_i\}$. In this complex representation, the matrix $\mathbb{T}_A$ is unitary, and $\mathbb{A}$ is Hermitian.

## Density matrix formalism

In this section we will depict the density matrix formalism in order to express and to demonstrate how the expectation value is obtained as it is formulated in quantum mechanics, from the point of view of probability theory. Recall that in the von Neumann formulation [16] of quantum theory, the expected value of an Hermitian operator $\hat{A}$ associated with an observable $A$ is given by the trace rule,

$$\langle \hat{A} \rangle := \text{Tr}\left(\hat{\rho}\hat{A}\right), \tag{17}$$

where $\hat{\rho}$ is a *density operator*. In our case, for an arbitrary state of knowledge $S$ we can write the expectation value of the measurement $A$ as

$$\left\langle R_A \right\rangle_S = \int d\lambda R_A(\lambda)P(\lambda|S) = \sum_{i=1}^{N} a_i P(a_i|S), \tag{18}$$

where in the last equality we have used $\int d\lambda R_A(\lambda)P(\lambda|a_i) = a_i$. Now, recognizing that the $\{a_i\}$ are the eigenvalues of the matrix $\mathbb{A}$, as given by Eq. (34), with corresponding eigenvectors $\boldsymbol{u}_i$, we can write them as $a_i = (\boldsymbol{u}_i)^T \mathbb{A} \boldsymbol{u}_i$ or, equivalently, we can use the spectral representation of $\mathbb{A}$, namely,

$$\mathbb{A} = \sum_{i=1}^{N} a_i \vec{u}_i (\vec{u}_i)^T.$$

which are known as quadratic forms in statistics [17].

Then, the expectation in state $S$ is

$$\left\langle R_A \right\rangle_S = \sum_{i=1}^{N} p_i(\boldsymbol{u}_i)^T \mathbb{A}\boldsymbol{u}_i = \sum_{l,m} \rho_{ml} A_{lm} = \mathrm{Tr}\,(\rho\mathbb{A}), \tag{19}$$

with $p_i = P(a_i|S)$. The density matrix $\rho$ associated to the state of knowledge $S$ is defined as

$$\rho = \sum_{i=1}^{N} p_i \boldsymbol{u}_i \boldsymbol{u}_i^T. \tag{20}$$

This is a properly defined density matrix because the $p_i$ are probabilities of discrete propositions, non-negative and adding up to 1. We can see that every system where we can write the expectation values of its properties in terms of the quadratic forms in Eq. (19) and leading to non-commutative operators, *can be considered as a fragile system.*

## Conclusions

We have shown that fragile systems with discrete properties can be analyzed in terms of genuine quantum theory, complete with non-commuting operators [18] and a density matrix formalism in complex Hilbert space. Since Bayesian degrees of belief are restricted by the rules of inference, and one of the aims of Basyesinism is the search for methods to translate information into probability assignments [19] while on the other hand any subjective state of knowledge about a quantum system can be summarized in a density operator $\rho$. In other words, the probabilities of measurement outcomes for a quantum system can be expressed as the trace of the product of a density operator and a projection operator corresponding to the measurement being performed.

Our derivation suggest the possibility of reducing most, if not all, of the content of quantum theory to the axiom: systems are perturbed when measured. This insight might be useful in developing necessary or sufficient conditions for violations of Bell's theorem in hidden-variable theories. As a matter of fact, the goal was to describe systems that are modified with the observation, in which quantum mechanics it is a particular case of this theory, in this sense our framework does not depend on the scale of the phenomena to be studied.

This not only gives a strong probabilistic justification for the fact that Nature itself seems to be described by quantum theory, but it also opens the possibility of justifying the use of the structure of quantum theory as an inference tool in problems involving fragile systems outside physics, in areas such as biology, data analysis [20], dynamical systems [11] among others. For instance, it would be possible to apply this tool to biological models, under the perspective of autopoietic systems [21] having self-modifying properties. There is also the interesting possibility of applying our results as a formal justification of the recent idea of quantum cognition [8] in which the object of study is human logic and human decisions, as well as in conductual psychology, where it is possible to modify the behavior of an individual by the observation of themselves.

## CRediT authorship contribution statement

**Yasmín Navarrete:** Writing – review & editing, Writing – original draft, Methodology, Investigation, Conceptualization. **Sergio Davis:** Writing – review & editing, Supervision, Methodology, Investigation, Formal analysis, Conceptualization.

## Declaration of competing interest

The authors declare the following financial interests/personal relationships which may be considered as potential competing interests: Yasmin Navarrete reports financial support was provided by National Agency for Research and Development. If there are other authors, they declare that they have no known competing financial interests or personal relationships that could have appeared to influence the work reported in this paper.

## Acknowledgments

The authors gratefully acknowledge funding from ANID FONDECYT 1220651.

## Appendix A Obtaining the linear transformation matrix $\mathbb{T}_A$

We introduce the basis functions $\{\phi_i\}$ into Eq. (7), so we can write Eq. (1) as

$$\sum_{j=1}^{n} w_j \phi_j(\lambda') = \int d\lambda \sum_{i=1}^{n} v_i \phi_i(\lambda) P(\lambda'|\lambda, A). \tag{21}$$

Multiplying both sides of the Eq. (21) by $\phi_k(\lambda')$ and integrating over $\lambda'$, we have

$$\sum_{j=1}^{n} w_j \int d\lambda' \phi_j(\lambda')\phi_k(\lambda') = \sum_{i=1}^{n} v_i \int d\lambda d\lambda' \phi_i(\lambda) P(\lambda'|\lambda, A)\phi_k(\lambda'). \tag{22}$$

Now, using the orthonormality condition (Eq. (9)) the left-hand side of Eq. (22) reduces to

$$\sum_{j=1}^{n} w_j \int d\lambda' \phi_j(\lambda')\phi_k(\lambda') = w_k, \tag{23}$$

which then can be written as follows,

$$w_k = \sum_{i=1}^{n} v_i \int d\lambda d\lambda' \phi_i(\lambda) P(\lambda'|\lambda, A)\phi_k(\lambda'). \tag{24}$$

Defining the matrix elements of $\mathbb{T}_A$ as

$$T_{ij}^{(A)} = \int d\lambda d\lambda' \phi_i(\lambda') P(\lambda'|\lambda, A)\phi_j(\lambda), \tag{25}$$

we can readily obtain the matrix equation corresponding to Eq. (1), namely

$$w_k = \sum_{i=1}^{n} T_{ki}^{(A)} v_i. \tag{26}$$

## Appendix B The eigenvalue equation for the matrix $\mathbb{A}$

For this we instead use the quantity:

$$Q(\lambda') := \int d\lambda R_A(\lambda) P(\lambda|S) P(\lambda'|\lambda, A), \tag{27}$$

such that for $S = a_k$, $P(\lambda|S) = P(\lambda|a_k) = g_k(\lambda)$. According to Eq. (4), $\lambda$ has zero probability if $R_A(\lambda) \neq a_k$. Considering this, we see that for $P(\lambda|S) = g_k(\lambda)$ it must hold that

$$Q(\lambda') = a_k \int d\lambda g_k(\lambda) P(\lambda'|\lambda, A) = a_k g_k(\lambda'), \tag{28}$$

based on the fixed points of the equation, where the second equality holds because of Eq. (5). Finally Eq. (27) yields

$$a_k g_k(\lambda') = \int d\lambda R_A(\lambda) g_k(\lambda) P(\lambda'|\lambda, A). \tag{29}$$

At this point we can write $g_k$ in terms of the basis $\{\phi_i\}$ on both sides, obtaining

$$a_k \sum_{l=1}^{n} u_l \phi_l(\lambda') = \sum_{j=1}^{n} u_j \int d\lambda R_A(\lambda)\phi_j(\lambda) P(\lambda'|\lambda, A). \tag{30}$$

where

$$g_k(\lambda) = \sum_{i=1}^{n} u_i \phi_i(\lambda). \tag{31}$$

Multiplying both sides by $\phi_i(\lambda')$ and integrating over $\lambda'$ we have

$$\int d\lambda' a_k \sum_{l=1}^{n} u_l \phi_l(\lambda')\phi_i(\lambda') = \int d\lambda' \sum_{j=1}^{n} u_j \int d\lambda R_A(\lambda)\phi_j(\lambda) P(\lambda'|\lambda, A)\phi_i(\lambda'), \tag{32}$$

therefore, the elements of the $A_{ij}$ are given by

$$A_{ij} = \int d\lambda d\lambda' R_A(\lambda)\phi_i(\lambda')P(\lambda'|\lambda, A)\phi_j(\lambda), \tag{33}$$

and after using the orthonormality on the left-hand side, we can finally arrive at the eigenvalue problem

$$\mathbb{A} \cdot \boldsymbol{u}_i = a_i \boldsymbol{u}_i. \tag{34}$$

## Data availability

No data was used for the research described in the article.